# A New Derivation of the Time-Dependent Schrödinger Equation from Wave and Matrix Mechanics

Luca Nanni*

Zambon SpA, Via della Chimica 9, I-36100 Vicenza, Italy

*E-mail of corresponding author: luca.nanni@student.unife.it

**Abstract**

An alternative method is proposed for deriving the time-dependent Schrödinger equation from the pictures of wave and matrix mechanics. The derivation is of a mixed classical–quantum character, since time is treated as a classical variable, thus avoiding any controversy over its meaning in quantum mechanics. The derivation method proposed in this paper requires no ad hoc assumption and avoids going through a second-order differential equation that can be reduced to the well-known time-dependent Schrödinger equation only postulating a complex wavefunction with a time dependence given by $exp(-iEt/\hbar)$, as did by Schrödinger in its original paper of 1926 [1].

## 1      Introduction

The time-dependent Schrödinger equation (TDSE) is widely used in quantum theory, especially in the study of physical phenomena for which the potential energy of the system explicitly depends on time (e.g., spectroscopic theory) [2]. Unlike the time-independent Schrödinger equation, which is easily derived from the classical wave equation and the de Broglie relation, the TDSE cannot be easily formulated by elementary methods and, in some textbooks, is simply introduced as an axiom. Schrödinger himself encountered difficulty in deriving the TDSE, initially aiming to obtain a quantum equation similar to the classical one for electromagnetic waves [1]. An exhaustive discussion of the principal methods of derivation is given by Briggs & Rost [3].
The explicit form of the TDSE is

$$i\hbar \frac{\partial}{\partial t}|\psi\rangle = H|\psi\rangle . \qquad (1)$$

In modern quantum mechanics text books only little mention is made on the classical nature of the time and on the fact that TDSE does not correspond to the energy conservation of the mechanical system. This suggests that it is worth deepening these fundamental aspects of quantum theory with the aim to obtain a more robust and logical derivation of TDSE. We prove below that the operator $i\hbar(\partial/\partial t)$ just represents the total energy of the system and can be easily derived in the frameworks of both wave and matrix mechanics (i.e., the Schrödinger and Heisenberg pictures). Then, using this result and the properties of commutative Hermitian operators [4], we prove that the TDSE can be derived without the need to assign a quantum operator to the time variable, thus avoiding a problem that is still a matter of debate and even controversy in quantum physics [5]. These steps are the key to formulate in a rational manner the TDSE, thus avoiding having to introduce it in the quantum theory as a postulate.

A new proof of the TDSE has already been provided recently by Sanayei [6] based on purely mathematical assumptions; here we start from physical concepts based on the parallels between geometric optics and Hamiltonian mechanics, making use of a simpler mathematical formalism. All that is required is basic knowledge of differential calculus and the algebra of quantum commutators.

## 2　　The Operator $i\hbar(\partial/\partial t)$

**2.1 Schrödinger Picture**

In many quantum mechanics textbooks, the wave theory of matter is introduced by highlighting the parallels between geometric optics and Hamiltonian dynamics [7,8,9]. More precisely, the latter is treated as a limiting case of the former. In geometric optics, the equation of a plane wave is given by

$$\boldsymbol{k} \cdot \boldsymbol{r} - \omega t = \text{const}, \tag{2}$$

while its phase velocity (which, in vacuum, coincides with the group velocity) is

$$v_f = v_g = \omega/k. \tag{3}$$

Let us consider a physical system whose motion is described by the vectors p and r given by

$$\boldsymbol{p} = (p_x(p_1, \dots, p_n), p_y(p_1, \dots, p_n), p_z(p_1, \dots, p_n)),$$

$$\boldsymbol{r} = (x(q_1, \dots, q_n), y(q, \dots, q), z(q_1, \dots, q_n)),$$

where $p_i$ and $q_i$ are the usual canonical variables. We suppose that the time dependence is limited to the position variables.

The equation of motion of a material particle that is acted upon by conservative forces, whose resultant is zero (steady state), is the following:

$$\boldsymbol{p} \cdot \boldsymbol{r} - \omega t = \text{const}. \tag{4}$$

This is similar to Eq. (4), but with the wavevector k replaced by the momentum p and the angular velocity ω replaced by the total energy E. By taking the time derivative of Eq. (4), we obtain

$$\frac{\partial}{\partial t}(\boldsymbol{k} \cdot \boldsymbol{r} - \omega t) = \dot{\boldsymbol{k}} \cdot \boldsymbol{r} + \boldsymbol{k} \cdot \dot{\boldsymbol{r}} - \omega = 0.$$

Supposing the wavevector to be time-independent (as was just done for the linear momentum vector), we obtain

$$\boldsymbol{k} \cdot \dot{\boldsymbol{r}} - \omega = 0 \quad \Rightarrow \quad \|\dot{\boldsymbol{r}}\| = r = \omega/k = v_g, \tag{5}$$

which is simply the phase velocity of the plane wave. Proceeding in the same way, we can take the time derivative of Eq. (5):

$$\frac{\partial}{\partial t}(\boldsymbol{p} \cdot \boldsymbol{r} - Et) = \dot{\boldsymbol{p}} \cdot \boldsymbol{r} + \boldsymbol{p} \cdot \dot{\boldsymbol{r}} - E = 0.$$

Since the resultant of the forces is zero, so is the time derivative of the linear momentum:

$$\boldsymbol{p} \cdot \dot{\boldsymbol{r}} - E = 0 \quad \Rightarrow \quad \|\dot{\boldsymbol{r}}\| = E/p = v, \tag{6}$$

where v is the velocity of the particle. Eqs. (5) and (6) are entirely analogous and summarize the parallelism between waves and particles.

We now turn to the quantum realm by invoking the third postulate of quantum mechanics, which states that every observable of a physical system is associated with a Hermitian operator with a complete set of eigenfunctions. The linear momentum and the particle position must then be replaced by the following operators:

$$\boldsymbol{p} \to i\hbar \nabla, \qquad \boldsymbol{r} \to \hat{r}.$$

Therefore, the operator form of Eq. (6) becomes

$$E = \boldsymbol{p} \cdot \frac{\partial \boldsymbol{r}}{\partial t}, \qquad \hat{E} = i\hbar \nabla \cdot \frac{\partial \hat{r}}{\partial t} = i\hbar \frac{\partial}{\partial t},$$

where the operator product $\nabla \cdot (\partial \hat{r}/\partial t)$ is given by

$$\nabla \cdot \frac{\partial \hat{r}}{\partial t} = \left( \frac{\partial}{\partial x}, \frac{\partial}{\partial y}, \frac{\partial}{\partial z} \right) \cdot \left( \sum_{i=1}^{n} \frac{\partial x}{\partial q_i} \frac{\partial q_i}{\partial t}, \sum_{i=1}^{n} \frac{\partial y}{\partial q_i} \frac{\partial q_i}{\partial t}, \sum_{i=1}^{n} \frac{\partial z}{\partial q_i} \frac{\partial q_i}{\partial t} \right).$$

We have proved that the operator $i\hbar(\partial/\partial t)$ represents the total energy of the quantum system. In doing so, we never needed to think of time as an operator.

### 2.2 Heisenberg Picture

We next show that the result obtained in the previous section can also be achieved in the Heisenberg picture (matrix mechanics) [10]. To this end, we first recall that in the Heisenberg picture the operators evolve over time, while in wave mechanics the time dependence is described by the wavefunction.

Consider Eq. (4), describing the same physical system (steady state); by taking its time derivative and supposing all quantities to be Heisenberg operators, we obtain

$$\hat{E} = \hat{p} \cdot \frac{\partial \hat{r}}{\partial t}. \tag{7}$$

Using the Heisenberg equations, the time derivative $\partial \hat{r}/\partial t$ of the position operator is given by

$$\frac{\partial \hat{r}}{\partial t} = \frac{1}{i\hbar} [\hat{r}, H],$$

where H is the usual Hamiltonian operator of the system. Substituting this derivative in Eq. (7) and recalling the explicit form of the momentum operator, we obtain

$$\hat{E} = \hat{p} \cdot \frac{\partial \hat{r}}{\partial t} = -i\hbar \frac{\partial}{\partial r} \frac{1}{i\hbar} [\hat{r}, H] = -\frac{\partial}{\partial r} [\hat{r}, H] = -\frac{\partial}{\partial r} (\hat{r}H - H\hat{r}).$$

Supposing that the Hamiltonian is the sum of an operator term representing the kinetic energy $K(p)$, which is a function of only the momenta, and an operator term representing the potential energy $U(r)$, which is a function only of the coordinates and does not explicitly depend on time, we have

$$\hat{E} = -\frac{\partial}{\partial r}(\hat{r}H - H\hat{r}) = -\frac{\partial}{\partial r}\{\hat{r}(\hat{K}(p) + \hat{U}(r)) - (\hat{K}(p) + \hat{U}(r))\hat{r}\}$$

$$= -\frac{\partial}{\partial r}\{\hat{r}\hat{K}(p) + \hat{r}\hat{U}(r) - \hat{K}(p)\hat{r} - \hat{U}(r)r\}$$

$$= -\hat{K}(p) - \hat{r}\frac{\partial \hat{K}}{\partial r} - \hat{U}(r) - \hat{r}\frac{\partial \hat{U}}{\partial r} + \frac{\partial \hat{K}}{\partial r}\hat{r} + \hat{K}(p) + \frac{\partial \hat{U}}{\partial r}\hat{r} + \hat{U}(r)$$

$$= \hat{K}(p) + \hat{U}(r) + \hat{r}\frac{\partial \hat{U}}{\partial r} - \hat{K}(p) - \frac{\partial \hat{U}}{\partial r}\hat{r} - \hat{U}(r)$$

$$= \left(H + \frac{\partial \hat{U}}{\partial r}\hat{r} - H - \hat{r}\frac{\partial \hat{U}}{\partial r}\right) = (-\hat{F}\hat{r} + \hat{r}\hat{F}) = (-\dot{\hat{p}}\hat{r} + \hat{r}\dot{\hat{p}}) = [\hat{r}, \dot{\hat{p}}],$$

where the terms in $\partial \hat{K}/\partial r$ vanish because the kinetic energy operator depends only on the linear momentum. Resuming, we have

$$\hat{E} = \hat{p} \cdot \frac{\partial \hat{r}}{\partial t} = [\hat{r}, \dot{\hat{p}}].$$

Now we have to work on the commutator $[\hat{r}, \dot{\hat{p}}]$ as follows:

$$[\hat{r}, \dot{\hat{p}}] = (\hat{r}\dot{\hat{p}} - \dot{\hat{p}}\hat{r}) = \left(\hat{r}\frac{\partial \hat{p}}{\partial t} - \frac{\partial \hat{p}}{\partial t}\hat{r}\right)$$

$$= \left\{\frac{\partial}{\partial t}(\hat{r}\hat{p}) - \hat{p}\frac{\partial \hat{r}}{\partial t} - \frac{\partial}{\partial t}(\hat{p}\hat{r}) + \frac{\partial \hat{r}}{\partial t}\hat{p}\right\} = \frac{\partial}{\partial t}[\hat{r}, \hat{p}] = i\hbar\frac{\partial}{\partial t},$$

where we have made use of the well-known commutation relation $[\hat{r}, \hat{p}] = i\hbar$. We have arrived at the same result as obtained in the Schrödinger picture: the total-energy operator of the system is given by $i\hbar(\partial/\partial t)$. For mathematical convenience, we have made use of a momentum operator of the form $p \rightarrow -i\hbar\nabla$ instead of $p \rightarrow i\hbar\nabla$; the two are completely equivalent because they are Hermitian.

Obtaining the same result as achieved in the Schrödinger picture is simply a consequence of the physical–mathematical equivalence of the two quantum theories (wave and matrix mechanics) [11].

## 3    Derivation of the Time-Dependent Schrödinger Equation

The operator $i\hbar(\partial/\partial t)$ is Hermitian because it is self-adjoint:

$$\left(i\hbar\frac{\partial}{\partial t}\right)^\dagger = (-i\hbar)\left(-\frac{\partial}{\partial t}\right) = i\hbar\frac{\partial}{\partial t}.$$

Moreover, if the Hamiltonian operator H does not depend explicitly on time, it commutes with $i\hbar(\partial/\partial t)$:

$$\left[i\hbar\frac{\partial}{\partial t}, H\right] = 0.$$

Therefore, the two operators have the same eigenspace $\{|\psi_i\rangle\}$ [4], and since both represent the same observable, it is logical to immediately conclude that they have the same spectrum of eigenvalues. However, we provide a formal proof of this statement.

Consider the following two eigenvalue equations:

$$i\hbar \frac{\partial}{\partial t}|\psi_i\rangle = \varepsilon_i|\psi_i\rangle, \qquad H|\psi_i\rangle = \varepsilon'_i|\psi_i\rangle,$$

where the eigenvalues are real and nonzero. Since the Hamiltonian does not explicitly depend on time, it follows that we can then write

$$i\hbar \frac{\partial}{\partial t}H|\psi_i\rangle = i\hbar \frac{\partial}{\partial t}\varepsilon'_i|\psi_i\rangle, \qquad Hi\hbar \frac{\partial}{\partial t}|\psi_i\rangle = H\varepsilon_i|\psi_i\rangle.$$

Subtracting the second equation from the first, we obtain

$$\left[i\hbar \frac{\partial}{\partial t}H - Hi\hbar \frac{\partial}{\partial t}\right]|\psi_i\rangle = \left[i\hbar \frac{\partial}{\partial t}\varepsilon'_i - H\varepsilon_i\right]|\psi_i\rangle,$$

and since the two operators commute, the left-hand side of this equality is zero:

$$\left[i\hbar \frac{\partial}{\partial t}\varepsilon'_i - H\varepsilon_i\right]|\psi_i\rangle = 0 \quad \Rightarrow \quad i\hbar \frac{\partial}{\partial t}\varepsilon'_i = H\varepsilon_i. \tag{8}$$

Multiplying both sides first by $\varepsilon'_i$ and then by $\varepsilon_i$, we obtain the following equalities:

$$\varepsilon'^2_i i\hbar \frac{\partial}{\partial t} = \varepsilon'_i\varepsilon_i H, \qquad \varepsilon_i\varepsilon'_i i\hbar \frac{\partial}{\partial t} = \varepsilon_i^2 H.$$

These equations are identical only if $\varepsilon_i\varepsilon'_i = \varepsilon'^2_i = \varepsilon_i^2$, which means that $\varepsilon_i = \varepsilon'_i$. Taking into account Eq. (8), we have

$$\left[i\hbar \frac{\partial}{\partial t}\varepsilon'_i - H\varepsilon_i\right]|\psi_i\rangle = i\hbar \frac{\partial}{\partial t}\varepsilon_i|\psi_i\rangle - H\varepsilon_i|\psi_i\rangle = 0,$$

from which we obtain the TDSE:

$$i\hbar \frac{\partial}{\partial t}|\psi_i\rangle = H|\psi_i\rangle.$$

If the Hamiltonian H has an explicit time dependence, it no longer commutes with $i\hbar(\partial/\partial t)$. In this case, we can construct a new time-independent operator by performing a change of variables in the classical Hamiltonian function $H_c = H(p, r, t)$ [6]:

$$q'_i = \begin{cases} q_i, & \text{if } i \leq n, \\ t, & \text{if } i = 0, \end{cases} \qquad p'_i = \begin{cases} p_i, & \text{if } i \leq n, \\ -\varepsilon, & \text{if } i = 0, \end{cases}$$

where $\varepsilon$ and t are two new conjugate variables. The transformed Hamiltonian is then given by

$$H'_c(p', r') = H(p, r, t) - \varepsilon, \tag{9}$$

which no longer has an explicit time dependence. The associated equations of motion in terms of the new variables are

$$-\frac{\partial H'_c(p', r')}{\partial q_i} = \dot{p}_i, \qquad \frac{\partial H'_c(p', r')}{\partial p_i} = \dot{q}_i. \tag{10}$$

For integers $1 \leq i \leq n$, these new equations are identical to the old ones, while for the new conjugate variables $q_0$ and $p_0$, they become

$$-\frac{\partial H'_c(p', r')}{\partial q_0} = -\frac{\partial H_c(p, r, t)}{\partial t} = \dot{p}_0 = -\varepsilon,$$
$$\frac{\partial H'_c(p', r')}{\partial p_0} = \frac{\partial H_c(p, r, t)}{\partial(-\varepsilon)} = \dot{t} = 1. \quad (11)$$

Again using the third postulate, we obtain the new Hamiltonian operator H′, thus returning to the original condition (the energy operator does not explicitly depend on time).

The derivation of TDSE presented in this paper is based on the assumption that the time is a parameter, preserving thus its classical nature. This is true when the external perturbations made on the quantum system are considered classically, just as in the spectroscopic theory [1]. Future researches must be developed in the case the quantum system and its environment are entangled and the time cannot be more considered classically.